\documentclass[traditabstract]{aa}   %%%%%%%%%%%% journal format 
\usepackage{graphicx}
\usepackage{txfonts}
\usepackage{color}
\usepackage{natbib}
\usepackage{xcolor}
\usepackage{epsfig}
\usepackage{hyperref}
\usepackage{longtable}
\usepackage[switch]{lineno}
\begin{document}
   \title{Jet outflow and gamma-ray emission correlations in S5 0716$+$714}

   \subtitle{ }

   \author{
          B. Rani \inst{1}
          \and T. P.\ Krichbaum \inst{1}
          \and A. P. Marscher \inst{2}
          \and S. G. Jorstad \inst{2}
          \and J. A. Hodgson \inst{1}
          \and L. Fuhrmann \inst{1}
          \and J. A. Zensus \inst{1} 
          }

   \institute{ 
              Max-Planck-Institut f{\"u}r Radioastronomie (MPIfR), Auf dem H{\"u}gel 69, D-53121 Bonn, Germany 
          \and 
            Institute for Astrophysical Research, Boston University, 725 Commonwealth Avenue, Boston, MA 02215, USA
           }

   \date{Received ---------; accepted ----------}

% \abstract{}{}{}{}{} 
% 5 {} token are mandatory
 
  \abstract
{ Using millimeter-very long baseline interferometry (VLBI) observations of the BL Lac object S5 0716+714 
from August 2008 to September 2013, we investigate variations in the core flux density and orientation of the sub-parsec 
scale jet i.e.\ position angle. The $\gamma$-ray data obtained by the {\it Fermi}-LAT (Large Area Telescope) 
are used to investigate the high-energy flux variations over the same time period. For the first time in any blazar, 
we report a significant correlation between the $\gamma$-ray flux variations and the position angle (PA) variations in 
the VLBI jet. The cross-correlation analysis also indicates a positive correlation such that the mm-VLBI core flux density 
variations are delayed with respect to the $\gamma$-ray flux by 82$\pm$32 days.   
This suggests that the high-energy emission is coming from a region located  $\geq$(3.8$\pm$1.9)~parsecs upstream of the 
mm-VLBI core (closer to the central black hole). These results imply that the observed inner 
jet morphology has a strong connection with the observed $\gamma$-ray flares. }

%___________________________________________________________


\maketitle

\section{Introduction}
The origin of high-energy emission has long been a key question in AGN (Active Galactic Nuclei) 
physics. A combination of high-resolution very long baseline interferometry (VLBI) images with 
broadband flux variability measurements is a unique way to probe the emission mechanisms at the bases of 
jets in a subclass of AGN called blazars, one of whose jets points towards the observer's line 
of sight. High-resolution mm-VLBI observations offer a unique possibility of studying the 
structural evolution in the inner parsecs of jets, which are proposed to be the sites of 
the high-energy emission region \citep[e.g.][and references therein]{lahteenmaki2003, 
rani2013_3c273, rani2013b, marscher2008, schinzel2012, lars2014}.

Non-radial motion, helical paths of the jet features, curved jet structures, and variations in the direction 
of the inner jet flow i.e.\ position angle or jet wobbling have often been 
observed in blazars \citep{britzen2009, bach2005, rastorgueva2011, lister2013, molina2014}. 
In some cases the inner jet position angle (PA) variations were found to 
correlate with the flux density 
variations at radio frequencies \citep{britzen2009, liu2012} and also with X-rays \citep{chatterjee2008}. 
However, a correlation between $\gamma$-ray flux and PA variations has not been reported so far.

In this paper, we report a correlation between $\gamma$-ray flux variations with 
the VLBI core flux density and the direction of the inner jet flow in the BL Lac object S5 0716+714. The mm-VLBI 
observations over the past five years were used to investigate the correlation. 
The aim of the study is to provide better constraints on the location of the high-energy emission region with 
an emphasis on the inner jet region kinematics and its 
correlation with the high-energy flux emission. A detailed analysis of the whole jet kinematics 
will follow in a subsequent paper.

The BL Lac object S5 0716+714 is one of the most intensively studied blazars because of its extreme variability 
properties across the whole electromagnetic spectrum \citep[][and references therein]{villata2008, fuhrmann2008, rani2013a, rani2013b, 
larinov2013}. The source has a featureless optical continuum with the redshift roughly 
constrained to the range 0.2315 $<$ z $<$ 0.3407 \citep{danforth2013}; here we used $z$$\approx$0.31 \citep{nilsson2008}. 
VLBI studies of the 
source show a core-dominated jet pointing towards the north 
\citep{bach2005, britzen2009}, and VLA (Very Large Array) observations show a halo-like jet misaligned with it by $\sim$90$^{\circ}$ \citep{antonucci1986}.
The broadband flaring behavior of the source is even more complex. The observed flux density light curves very often reflect 
rapid flaring activity superimposed on top of a broad and slow variability trend \citep{rani2013a, raiteri2003}.

%__________________________________________________________________

\section{Observations and data reduction}
\subsection{Very Long Baseline Interferometry}
For the jet kinematics study, we used the mm-VLBI data of the source observed between August 2008 and 
September 2013.  The 7~mm (43~GHz) data were a result of the Boston University group\footnote{http://www.bu.edu/blazars} monthly 
monitoring program with the Very Long Baseline Array (VLBA). The 3~mm (86~GHz) observations were performed using the Global mm-VLBI Array 
(GMVA\footnote{http://www3.mpifr-bonn.mpg.de/div/vlbi/globalmm/index.html}). In total, we had 
observations at 63 epochs over the past 5 years. 
Data reduction was performed using the standard tasks of the Astronomical Image Processing System 
(AIPS) and Difmap \citep{shepherd1997}. 
Imaging of the source (including amplitude and phase self-calibration) was performed using the {\it CLEAN} 
algorithm \citep{hogbom1974} and {\it SELFCAL} procedures in the Difmap package \citep{shepherd1997}. 
Further details of the data reduction can be found in \citet{jorstad2005}.

We modeled the observed brightness distribution of the radio emission by multiple circular Gaussian 
components providing positions, flux densities, and sizes of the distinct bright features in the jet using the 
Difmap package. For all model fits, we used the brightest component as a reference and fixed
its position to (0, 0). The final number of jet components necessary to fit the data were adequately 
achieved when adding an extra component did not lead to a significant improvement in the fit. 
Uncertainties of the parameter fits were estimated following \citet{krichbaum1998} and \citet{jorstad2005}.

  \begin{figure}
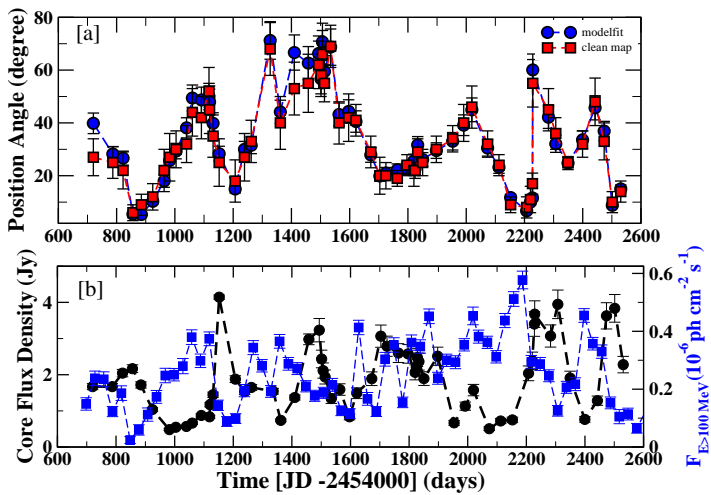

\includegraphics[scale=0.35,angle=0, trim=2 23 0 0, clip]{time_PA.eps}
\includegraphics[scale=0.35,angle=0, trim=-7 0 0 0, clip]{Gamma_ray_core.eps}
 \caption{(a): Position angle (PA) variations in the central region of the jet. The blue circles show the 
PA calculated using model-fitting, while those estimated directly from clean maps are in red (square symbols). 
(b): mm-VLBI core flux density variations (black circles) over the same time period superimposed with the monthly averaged 
$\gamma$-ray flux variations (blue squares).   }
\label{plot_fig1}
\end{figure}

\subsection{Gamma rays} 
We employed here the 100~MeV -- 300~GeV data of the source from August 04, 2008 to 
September 30, 2013, which were observed in survey mode by the {\it Fermi}-LAT 
\citep[Large Area Telescope,][]{atwood2009}. We analyzed the LAT 
data using the standard ScienceTools (software version v9.32.5) and instrument response function P7REP$\_$SOURCE. 
Photons in the Source event class were selected for the analysis.  As the VLBI observations of the 
source have an average sampling of $\sim$1 month, we preferred a time binning of one month for the 
$\gamma$-ray photon flux light curves. We obtained similar results for weekly binned light curves. 
The monthly binned light curves of the 
source at E$>$100~MeV were produced by modeling the spectra over each bin by a simple power law 
(N(E) = N$_0$ E$^{-\Gamma}$, N$_0$ : prefactor, and $\Gamma$ : power law index). For this analysis, we 
used the unbinned maximum-likelihood algorithm \citep{mattox1996}. 
The analysis performed in this paper is very 
similar to that reported in \citet{rani2013a}, to which we refer for details.

\section{Analysis and results}

\subsection{Jet orientation variations}
\label{pa_var} 
To determine the inner jet orientation, we used an annular region up to 0.2~mas 
from the core at (0,0).  We fitted a straight line between 
the core and its adjacent component (see Fig.\ A1). The PA of this 
line provided a reasonably good estimate of the direction of the inner portion of the jet. However, it 
should be noted that the derived position 
angle values could be influenced by choices made for the fitted Gaussians (e.g.\ elliptical vs.\ circular) and the total 
number of Gaussian components in the inner jet region. We therefore explored an independent approach for calculating 
the inner jet PA. We determined the inner jet PA at each epoch by taking a flux density-weighted
PA average of all the clean delta components\footnote{The clean delta 
components represent the de-convolved brightness distribution of the source structure} three times above the image noise level in the 
annular region (see Fig.\ A1). We used the 1~$\sigma$ distribution of the clean delta components around the mean 
PA axis as uncertainties. The two approaches suggest consistent PA variations  as shown in Fig.\ \ref{plot_fig1} (a).

  \begin{figure}
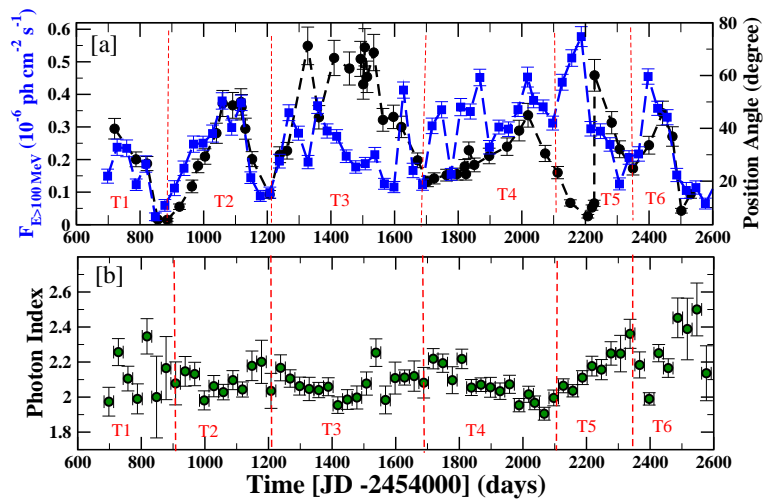

 \includegraphics[scale=0.39,angle=0, trim=0 21 0 0, clip]{Gamma_ray_flux_PA.eps}
 \includegraphics[scale=0.391,angle=0, trim=-15 0 1 0, clip]{Gamma_indx.eps}
 \caption{(a) : The monthly averaged $\gamma$-ray flux light curve (blue squares) superimposed on top of the 
model-fitted PA curve (black circles). (b) : Gamma-ray photon index versus time.       }
\label{plot_fig2}
\end{figure}

  \begin{figure}
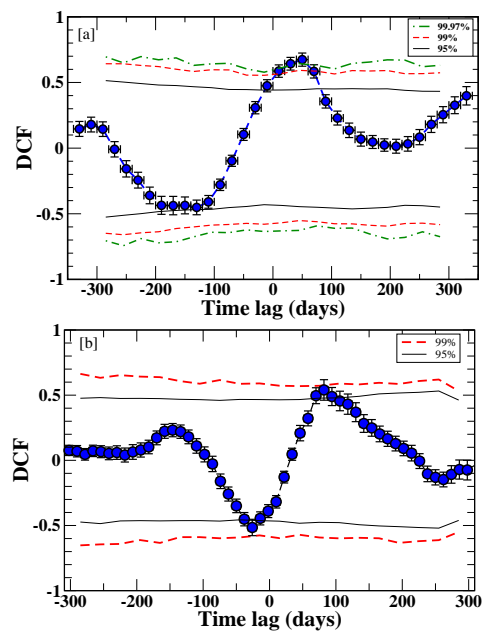

   \centering
 \includegraphics[scale=0.25,angle=0, trim=0 0 0 0, clip]{Gamma_PA_dcf.eps}
 \includegraphics[scale=0.25,angle=0, trim=0 0 0 0, clip]{Gamma_core_dcf.eps}
 \caption{(a): DCF analysis curve of the $\gamma$-ray flux and PA variations in the inner jet region.  
The lines show confidence levels as discussed in the text. (b): DCF analysis curve between the $\gamma$-ray and core 
flux density variations. A positive time lag implies that $\gamma$-ray flux variations lead core flux density and 
PA variations. }
\label{plot_fig3}
\end{figure}

\subsection{Gamma-ray flux vs.\ PA variations}
\label{Gamma_PA}
To compare the $\gamma$-ray flux variations with the PA variations, we plot the two 
on top of each other (see Fig.\ \ref{plot_fig2} a). For a visual comparison, we divide  the whole period into six 
segments, T1 to T6. The inner-jet PA and $\gamma$-ray  flux appear to vary together during T1 and T2. After this, the 
variations diverge until T6, when they are again similar. This  suggests a relation between the two properties 
of the blazar that is sometimes strong and at other times complex or non-existent. 

To quantify the apparent correlation of the PA and $\gamma$-ray flux variations, we employed the discrete 
cross-correlation function (DCF) analysis method \citep{edelson1988}. Figure \ref{plot_fig3} (a) shows the DCF analysis results of
the monthly averaged $\gamma$-ray flux versus the inner jet PA.  The DCF curve shows a prominent peak at 
47$\pm$22 days.  The given uncertainty in the time lag value 
here and in the following Sections is the half-width at 
the 90$\%$ point of the best-fit Gaussian function to the DCF curve. The significance of the DCF analysis was 
tested using simulations as discussed in Appendix \ref{DCF_sig}, and the 95, 99 and 99.97$\%$ confidence levels 
are shown in Fig.\ \ref{plot_fig3} (a).  The simulations imply that the significance of the 
measured correlation is $>$3$\sigma$. It is important to note that the PA values are 
measured by fixing the core position to (0,0) within a region up to 0.2 mas; consequently, the observed PA 
variations can either be related to the mm-VLBI core and/or the jet flow farther downstream of the core.
The observed time lag therefore cannot be used for physical calculations, e.g., to estimate the distance between 
the two emission regions, as the reference/zero point of the PA measurements is arbitrary; but, importantly the 
$\gamma$-ray flux variations correlate significantly with PA variations of the inner jet.

 In Fig.\ \ref{plot_fig2} (b), we plot the $\gamma$-ray photon index ($\Gamma$) as a function of 
time.  The main deviations from apparently random fluctuations are seen during T4 and T5.  During T4, 
$\Gamma$  drops from 2.19$\pm$0.05 to 1.90$\pm$0.04, and later the spectrum softens to a photon 
index value = 2.36$\pm$0.08 at the end of T5. 
Since the photon index variations cannot be interpreted as 
purely geometrical effects, this suggests that the flux variations are related to changes 
in both the physical conditions in the plasma and  Doppler beaming.

\subsection{Gamma-ray flux vs. core flux density variations}
\label{vlbi_flare_corr}
Figure \ref{plot_fig1} (b) shows the monthly binned $\gamma$-ray flux light curve plotted on 
top of the mm-VLBI core flux density\footnote{We used a mean value 
of the measured optically thin spectral index, $\alpha_{thin}$ = 0.4 \citep{rani2013a} to scale the 3~mm 
flux density measurements to those at 7~mm.}  light curve. 
For several events, the peak of the $\gamma$-ray flare appears to coincide with the onset of the radio flare, and 
to investigate it, we used the DCF method. The DCF curve (Fig.\ \ref{plot_fig3} b) shows a peak at 
82$\pm$32 days and a dip at $-$(29$\pm$25) days. The significance of the cross-correlation was tested 
via simulations as discussed in Section \ref{DCF_sig}. In Fig.\ \ref{plot_fig3} (b), the lines show the 
95 and 99$\%$ confidence levels. The simulations, therefore, revealed that the significance of 
both the correlation (at 82$\pm$32 days) and the anti-correlation (at $-$(29$\pm$25) days) is 
$\sim$99$\%$. An anti-correlation implies that $\gamma$-ray flares lead those at radio wavelengths or 
vice-versa such that the maximum of one coincides with the minimum of other. We found 
that the radio jet flux density further downstream 
of the core is very faint and does not show any correlated variation with the $\gamma$-ray light curve, suggesting that 
the $\gamma$-ray flares are not produced downstream of the core. This implies that 
$\gamma$-rays are produced upstream of the core, which is also supported by the positive 
correlation between $\gamma$-ray and core flux density light curves with the former leading the latter 
by 82$\pm$32~days.  We note that for a larger sample of $\gamma$-ray blazars, similar results ($\gamma$-ray 
leading mm-radio flares) were recently reported by \citet{lars2014}.

  \begin{figure}[h]
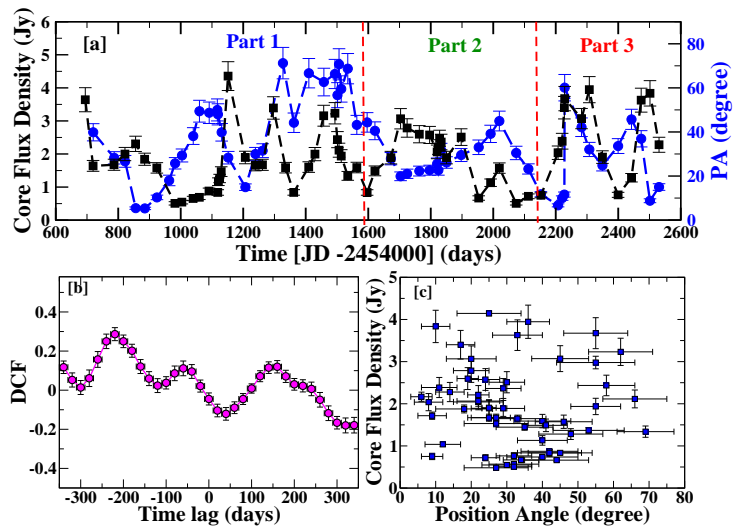

\includegraphics[scale=0.383,angle=0, trim=-2 0 0 0, clip]{core_flux_PA.eps}
\includegraphics[scale=0.1855,angle=0, trim=0 0 0 0, clip]{dcf_core_PA_new.eps}
\includegraphics[scale=0.1855,angle=0, trim=0 0 0 0, clip]{Core_PA_lin.eps}
 \caption{(a):  Core flux density light curve superimposed on the PA curve. (b): 
The DCF analysis curve of the core flux density vs.\ PA variations. (c): core flux density  vs.\ PA plot.        }
\label{plot_fig5}
\end{figure}

\subsection{Core flux density vs. PA variations}
\label{vlbi_core_PA}
 A comparison of the observed core flux density and PA variations is shown in Fig.\ \ref{plot_fig5} (a). 
For a visual comparison, we divide the whole period into three segments, Part 1 to 3. The core flux density 
and PA appear to vary together for Part 3, while the variations diverge during 
Part 1 and 2. The formal DCF analysis of all data does not reveal a significant correlation between 
the two (see Fig.\ \ref{plot_fig5} b). However, the absence of a significant correlation does not rule 
out a weak correlation, or much more complicated behavior. The core flux density vs.\ PA plot 
(Fig.\ \ref{plot_fig5} c) shows a ring-like pattern with its center at $\sim$40$^{\circ}$, which 
suggests some sort of correlation between the two.

\section{Discussion and Conclusion}
\label{discussion}
 Our analysis suggests a strong correlation between high-energy emission and inner jet morphology. We found 
a strong correlation between $\gamma$-ray flux variations and PA variations. The observed time lag of 82$\pm$32 
days between $\gamma$-ray and core flux density variations  places the $\gamma$-ray emission 
region  upstream of the mm-VLBI core by $\geq$(3.8$\pm$1.9)~parsec \citep[deprojected using a viewing 
angle, $\theta \leq 4.9^{\circ}$ and apparent jet speed $\beta_{apparent}$=10,][]{bach2005}. 
These correlations indicate that first some change in the jet structure 
triggers a $\gamma$-ray flare, and later the event has traveled 0.2 mas down the jet, so that either 
the jet PA between the core and the first jet component changes, or at that point the event has just traveled to 
the core and caused the core to shift in the transverse direction. In a simple scenario, we would also expect 
a correlation between core flux density  and PA variations in the optically-thin case; however the correlation 
could be weaker for a partially optically-thick core.

 Systematic variations in the orientation of 
parsec to sub-parsec scale VLBI jets have been observed in many
sources \citep[e.g.][and references therein]{bach2005, lister2013, molina2014}. The exact origin of these 
variations is not yet clear, although accretion disk precession, orbital motion of 
the accretion system, or instabilities (Magnetohydrodynamic (MHD), and/or  
Kelvin-Helmholtz (KH)) in the jet flow have all been suggested.
The relatively short variability time scales ($\sim$200~days) involved and the
observed non-ballistic jet motion in S5 0716+714 \citep{bach2005, britzen2009} most likely excludes geometric
precession due to a binary black hole.
In relativistic jet models \citep{blandford1982, blandford1977}, spatially bent jets and helical 
fluid patterns are a natural consequence of KH-instabilities \citep{hardee2011} and/or MHD
instabilities \citep{meier2001}. In such `magnetic jets', the helical structure is formed by a twist of 
the magnetic filaments through and around the conical jet, which may explain the observed PA variations 
in the source.

\begin{figure}
   \centering
\includegraphics[scale=0.35,angle=0, trim=75 62 0 40, clip]{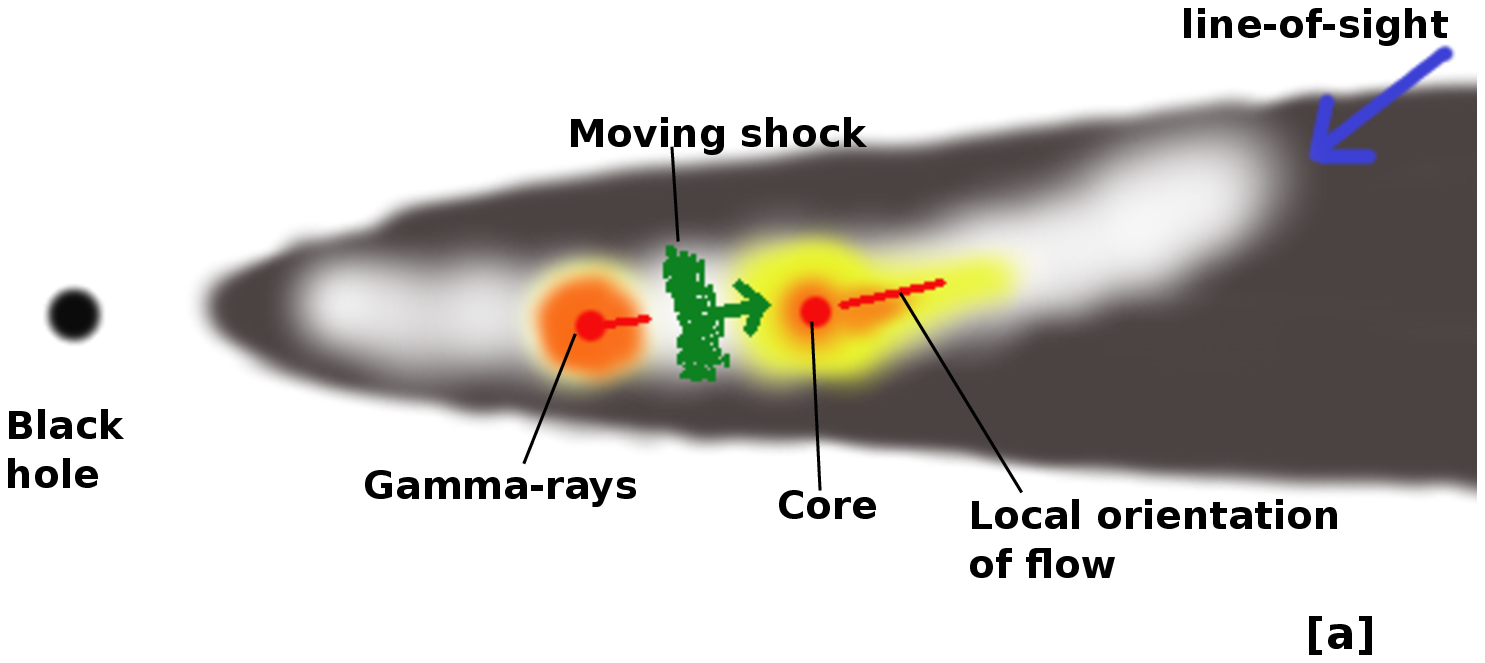}
\includegraphics[scale=0.35,angle=0, trim=75 70 0 40, clip]{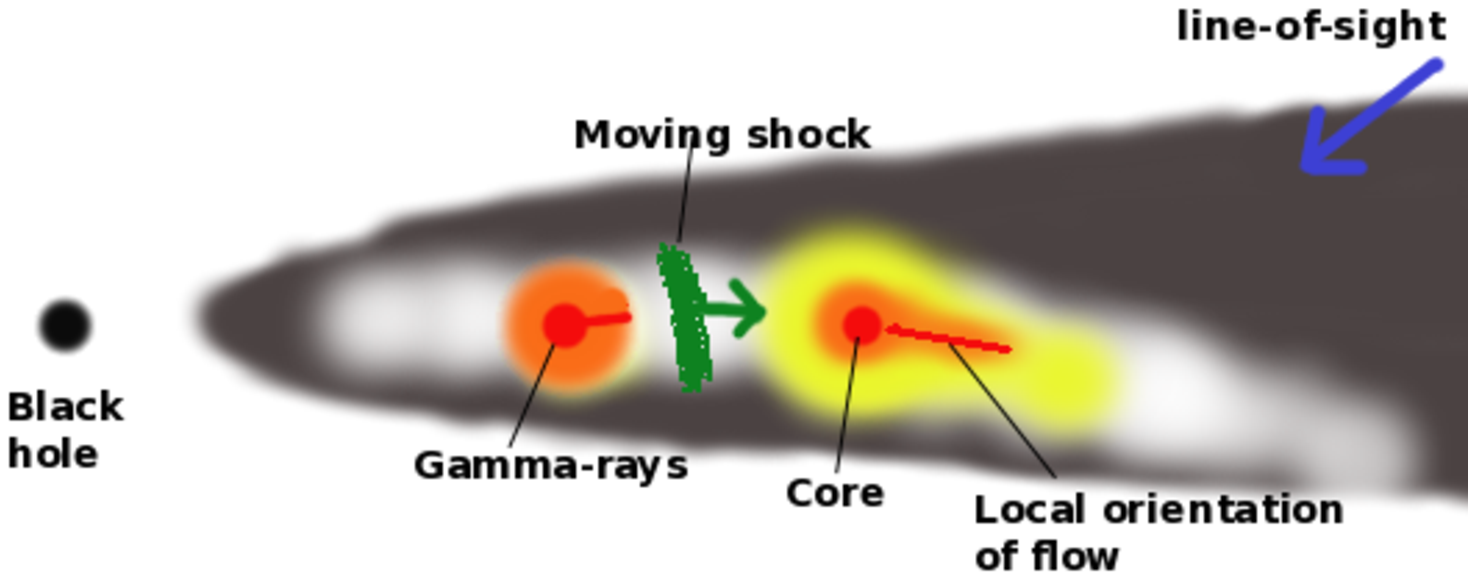}
 \caption{A sketch for the proposed scenario in the BL Lac S5 0716+714 (not to scale). 
The high density/pressure regions, shown in light gray color (superimposed on top of 
the underlying jet flow, which is in dark color), brighten relative to other regions of the jet by the 
passage of a moving shock. (a) : case for a strong correlation, and (b) : a weak correlation.        }
\label{plot_fig6}
\end{figure}

 In a tentative model (shown in Fig.\ \ref{plot_fig6}), the observed $\gamma$-ray flux and PA variations can be interpreted as 
a moving shock propagating down a relativistic jet with non-axisymmetric pressure and/or density gradients/patterns 
or a shock moving in a bent jet. A moving shock will induce significantly increased emission 
at the locations where it intersects with regions of enhanced electron density and/or magnetic 
field. The measured correlations suggest that the $\gamma$-ray flares precede the mm-VLBI core flares, and 
the time lag depends on the physical conditions of the emission region. Longer time lags can be expected 
via opacity effects and/or if the 
two emission regions are separated (as shown in Fig.\ \ref{plot_fig6}). Because Doppler boosting is 
a sensitive function of viewing angle, substantial changes in amplitude of jet emission can be seen 
by the observer. Correlated variations between the $\gamma$-ray emission and orientation of the jet flow is obvious 
if the two share the same boosting cone as shown in Fig.\ \ref{plot_fig6} (a). However if the two emission 
regions are pointed in different directions the correlation between $\gamma$-ray flux and PA would be 
weaker (Fig.\ \ref{plot_fig6} b). Therefore, this scenario successfully explains why sometimes we see 
very strong correlations and sometimes not (see Fig.\ \ref{plot_fig2}).  One could also consider 
instability patterns moving downstream passing the two emission regions at different angles, or even a rotation of 
the (helical) jet around its own z-axis. All models would cause very similar variations of the viewing angle, which is 
responsible for the observed correlation between $\gamma$-ray flux and 
apparent jet position angle.  A correlated variation between the core flux density and the PA curves is expected in 
a simple scenario. The expected correlation is however not detected with the current observations.

The observed correlation between the $\gamma$-ray flux variations and the inner jet kinematics are a 
challenge for the available relativistic jet models.  Sub-mm (3~mm/1~mm) VLBI monitoring with denser 
time sampling would be required to understand the parsec scale jet morphology in a better way. In addition 
to this, the aforementioned hypotheses need to be developed via magnetohydrodynamic simulations to 
provide a better understanding of the jet launching region.

\begin{acknowledgements}
The $Fermi$-LAT Collaboration acknowledges support from a number of agencies and institutes for both 
development and the operation of the LAT as well as scientific data analysis. These include NASA and 
DOE in the United States, CEA/Irfu and IN2P3/CNRS in France, ASI and INFN in Italy, MEXT, KEK, and JAXA 
in Japan, and the K.~A.~Wallenberg Foundation, the Swedish Research Council and the National Space Board 
in Sweden. Additional support from INAF in Italy and CNES in France for science analysis during the 
operations phase is also gratefully acknowledged. This study makes use of 43 GHz VLBA data from the 
VLBA-BU Blazar Monitoring Program (VLBA-BU-BLAZAR; http://www.bu.edu/blazars/VLBAproject.html), funded 
by NASA through the Fermi Guest Investigator Program. The VLBA is an instrument of the National Radio 
Astronomy Observatory. The National Radio Astronomy Observatory is a facility of the National Science 
Foundation operated by Associated Universities, Inc. We would like to 
thank Chuck Dermer, the internal referee for the Fermi-LAT Collaboration,  Eric C. Brown, 
Andrei Lobanov, Jeremy Perkins, and David Thompson for their useful suggestions 
and comments. BR is thankful to Moritz Boeck, S. Vaughan, D. Emmanoulopoulos, and I. McHardy for the 
useful discussion on calculating the significance of correlations. We thank the referee for 
constructive comments. 

\end{acknowledgements}

%\bibliographystyle{aa} % style aa.bst
%\bibliography{references} % your references Yourfile.bib

\begin{appendix}

\section{Testing the correlation significance}
\label{DCF_sig}
The significance of correlations in the DCF analysis was determined using simulations. 
To do so, we first estimated the PSD (power spectral density) slope of the $\gamma$-ray light 
curve following \citet{vaughan2005}. The estimated PSD slope for 0716+714 is $-$(0.93$\pm$0.24) . The next 
step is to generate the simulated light curves using the PSD slope. One can use the online available IDL code 
\footnote{http://astro.uni-tuebingen.de/software/idl/aitlib/timing/timmerlc.html}. However, the simulated light 
curves using this code follow a Gaussian distribution, which is normally not the case for the observed light curves 
because a majority of them are burst-like events. This has to be taken into account while estimating the 
correlation significances \citep[see][for details]{dimiterious2013}. We therefore checked the 
distribution of the observed $\gamma$-ray light curve, and we found that the underlying distribution is similar 
to a Gaussian one. We generated a series of 100,000 light curves using the online available IDL code. 
The next step is to sample the simulated 
light curves at the same times and bin widths as the observations. We cross-correlated the simulated $\gamma$-ray 
light curves with the observed PA curve. Finally, we calculated the distribution of the DCF values as a function 
of time lag. For each time delay, we estimated the 0.025 and 0.975 quantiles corresponding to the upper and lower
limits of the 95$\%$ confidence bands. However, these confidence levels are obtained for a single trial i.e.\ if 
we already know the time lag between the two light curves. Since we do not have the time lag information a priori, 
we always use a search window depending on the duration of observations. To correct for this effect 
(called the ``look elsewhere effect"), one has to include the total number of trials, which is the number of data points 
in the given search window \citep[for details see Section 4.1 in][]{vaughan2005}. For N points, the confidence level, $\epsilon$,   
is (1 $-$ (1 $-$ $\epsilon$)/N)$\times$100~$\%$ i.e.\ the 95$\%$ confidence level for 10 
data points in a given time lag range should be [1 $-$ (1 $-$ 0.95)/10]$\times$100 = 99.5$\%$ confidence level for a single data point.

 \begin{figure}[h]
 \includegraphics[scale=0.35,angle=0, trim=0 0 1 0, clip]{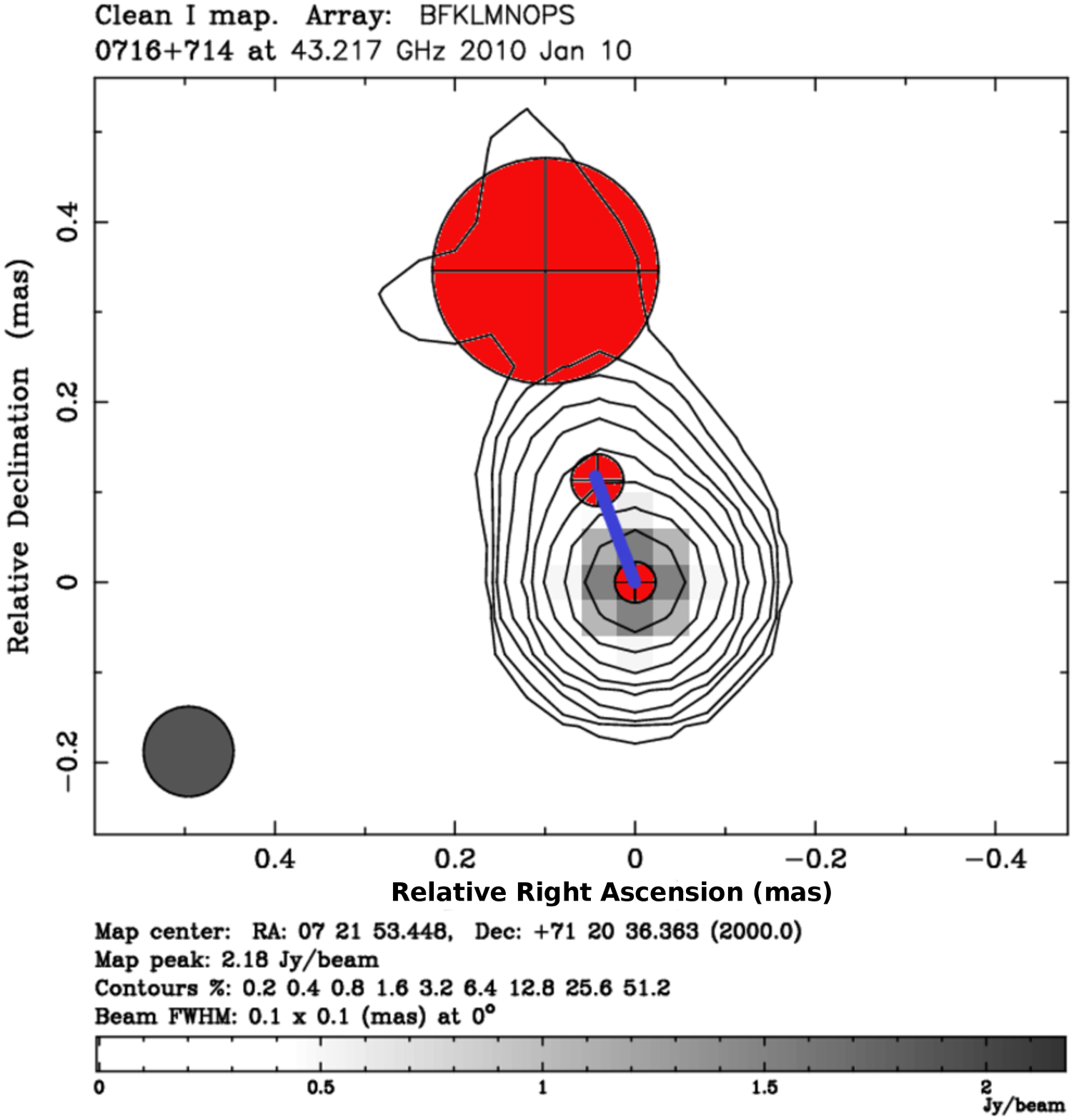} 
 \includegraphics[scale=0.35,angle=0, trim=0 0 0 0, clip]{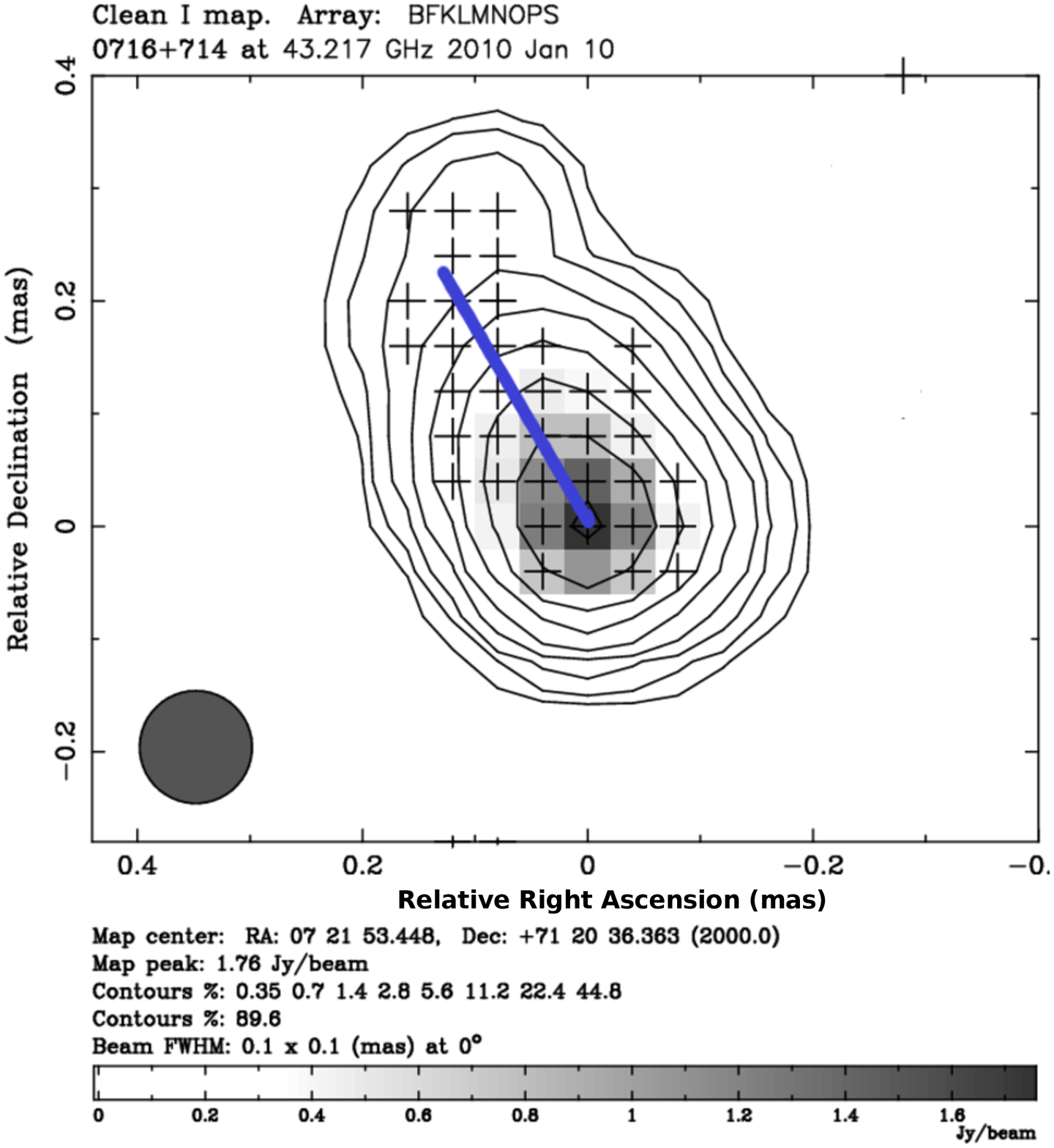}
 {Fig. A1. An example of 43~GHz VLBI images of S5 0716+714 observed on January 10, 2010 resolved with a 
beam size of 0.1~mas (gray circle in the bottom-left corner of maps). Contour maps are  superimposed with Gaussian model-fit 
components (red circles in the top panel) and clean delta 
components (plus symbols in the bottom panel).  The blue lines mark the inner jet orientation.     }
\label{plot_fig_images}
\end{figure}

\end{appendix}

\end{document}